\documentclass[pre, twocolumn]{revtex4-1}
\usepackage{amsmath,amssymb,bm,xcolor,mathrsfs,graphicx,subfigure,texdraw}
\usepackage{mathtools}

\usepackage{color}
\newcommand{\bq}{\mathbf{q}}
\newcommand{\br}{\mathbf{r}}
\newcommand{\SPY}{S_{PY}}
\newcommand{\lamPY}{\lambda_{PY}}
\newcommand{\rhom}{\overline{\rho}}

\begin{document}
\title{Jamming in hard sphere suspensions under shear}

\author{Noemi Barrera}
\email{noemibarrera@post.tau.ac.il}
\author{Moshe Schwartz}
\affiliation{School of Physics and Astronomy, Raymond and Beverly Sackler Faculty of Exact Sciences, Tel Aviv University}

\begin{abstract}
We consider a system of monodisperse hard spheres immersed in a sheared fluid. We obtain the distortion of the structure factor of the hard spheres at low shear rates, within a Percus-Yevick like framework. The consequent distortion of the pair distribution function is shown to affect the transition of the hard sphere fluid into a jammed state, which is similar to the transition to the state of random close packing in the absence of shear.
\end{abstract}
\maketitle

Systems of interacting extended objects subject to an imposed macroscopic flow have been attracting an ever increasing amount of interest over the last four decades.  Examples of such systems are emulsions, suspensions, colloids and foams. The imposed flow can be achieved by immersing the system in a sheared fluid or by shearing it directly, for example between moving plates. This kind of system has diverse applications in material fabrication, biological systems, paints, etc. and has been thoroughly investigated in several ways \cite{05Sti}. 

Most of the extended objects mentioned above are deformable but some of the experiments and numerical simulations have been done on model hard sphere systems \cite{93Koel, 80ClAc}. The present article deals with the effect of imposed flow on the correlations in a system of monodisperse hard spheres immersed in it. We are interested, in particular, in the distortion of the correlations and in shear-induced jamming due to macroscopic shear flow of the host fluid.\\

The distortion of the structure factor of the immersed objects in the presence of shear has been observed experimentally \cite{80ClAc} and explained theoretically \cite{84Ron}. A more interesting observation seen in simulations and experiments is shear-induced ordering that is accompanied, in some cases, by jamming of the particle system (\cite{88Ack}-\cite{97Farr}). \\
In this article, we are going to address the problem of jamming due to shear by relating it to random close packing in a system of hard spheres in equilibrium and to the distortion of the structure factor due to the shear of the host fluid.  \\

Let us start with an overdamped Langevin description of a system of $N$ identical particles contained in a cubic box of volume $V$ and periodic boundary conditions. The particles are  immersed in a host liquid flowing  with a prescribed velocity field $\mathbf{V}(\br)$,
\begin{equation}
\gamma\left[\dot{\br}_i-\mathbf{V}(\br_i)\right]=-\nabla\sum_{j=1}^N w(\br_i-\br_j)+\boldsymbol{\eta}_i(t),
\end{equation}                                   
where $\br_i$ is the radius vector of the $i$-th particle, $w(\br)$ is the two-body potential and the Cartesian components of the noise, $\eta_i^k$, obey $\langle \eta_i^k(t)\rangle=0$ and $\langle\eta_i^k(t)\eta_j^l(t')\rangle=2\sigma\delta_{ij}\delta_{kl}\delta(t-t')$. The Langevin equations above can be directly turned in the Fokker-Planck equation for the distribution of all the particles in the configuration space,
\begin{equation}\label{eqn:FP}
\dfrac{\partial P}{\partial t}=O\, P,
\end{equation}
where the operator $O$ has the following form 
\begin{equation}\nonumber
O=\dfrac{1}{\gamma}\left\{\sum_{i,\textrm{k}}\dfrac{\partial}{\partial r_{i\textrm{k}}}\left[kT\dfrac{\partial}{\partial r_{i\textrm{k}}}+\dfrac{\partial U}{\partial r_{i\textrm{k}}}-\gamma V_\textrm{k}(\br_i)\right]\right\},
\end{equation}
with $kT=\sigma/\gamma$ and 
\begin{equation}
U=1/2\sum_{i\neq j}w(|\br_i-\br_j|).
\end{equation} 
The Fokker Planck equation above can be transformed into an equation governing the evolution of the distribution $\hat{P}(\{\rho_\mathbf{q}\})$ of the collective degrees of freedom, which are the Fourier coefficients of the density $\rho(\br)=\sum_{i=1}^N \delta(\br-\br_i)$ \cite{02Edw}.\\Note that the Fourier coefficients are given in terms of the coordinates of the particles by
\begin{equation}
\rho_\mathbf{q}=\dfrac{1}{\sqrt{N}}\sum_i\exp(-i\mathbf{q}\cdot\mathbf{r}_i),
\end{equation}
so that $\rho_{\boldsymbol{0}}=\sqrt{N}$ is not a degree of freedom. \\
Since we are interested only in the steady state properties, we write down just the steady state equation $(\partial \hat{P}/\partial t=0)$,
\begin{equation}\label{eqn:steadystate}
\begin{split}
\dfrac{kT}{\gamma }\sum_{\mathbf{k}}&\left\{\textrm{k}^2\left[1+\dfrac{\rhom}{kT}w(\mathbf{k})+\rho_\mathbf{k}\dfrac{\partial}{\partial \rho_\mathbf{k}}\right]\right.-\nu \textrm{k}_z\dfrac{\partial \rho_\mathbf{k}}{\partial \textrm{k}_x}\dfrac{\partial}{\partial \rho_\mathbf{k}}+\\
&\;\qquad\qquad+N^{-1/2}\sum_{\mathbf{l}}\mathbf{k}\cdot\mathbf{l}\left[-\rho_{\mathbf{k}+\mathbf{l}}\dfrac{\partial^2}{\partial\rho_\mathbf{k}\partial\rho_{-\mathbf{k}}}+\right.\\
&\;\;\;\qquad\qquad\qquad+\left.\dfrac{\rhom}{kT}w(\mathbf{l})\left.\rho_{\mathbf{k}-\mathbf{l}}\rho_\mathbf{l}\dfrac{\partial}{\partial\rho_\mathbf{k}}\right]\right\}\hat{P}=0,
\end{split}
\end{equation} 
where $\rhom=N/V$, $w(\mathbf{k})$ is the Fourier transform of the two-body potential $w(\br)$ and $\nu=\gamma C/ kT$.\\

The equation above can be simplified by the random phase approximation (RPA), which consists of dropping the trilinear terms in the operator applied on the probability distribution in the equation above \cite{vinsc2003}.  Thus,
\begin{equation}\label{eqn:PhatSimpl}
\begin{split}
&\left\{\sum_\mathbf{k}\textrm{k}^2\left[1+\dfrac{\rhom}{kT}w(\mathbf{k})\right.\right.+\left(1+\dfrac{\rhom}{kT}w(\mathbf{k})\right)\rho_\mathbf{k}\dfrac{\partial}{\partial \rho_\mathbf{k}}+\\
&\qquad\qquad\left.\left. +\dfrac{\partial^2}{\partial\rho_\mathbf{k}\partial\rho_{-\mathbf{k}}}\right]-\nu \textrm{k}_z\dfrac{\partial\rho_\mathbf{k}}{\partial \textrm{k}_x}\dfrac{\partial}{\partial\rho_\mathbf{k}}\right\}\hat{P}=0,
\end{split}
\end{equation}
Multiplying Eq.\eqref{eqn:PhatSimpl} by $1/2\rho_\mathbf{q}\rho_{-\mathbf{q}}$ and integrating by parts over all the collective variables lead directly to a closed form equation for the structure factor $S(\bq)=\langle \rho_\bq\rho_{-\bq}\rangle$, 
\begin{equation}\label{eqn:mainSHEAR}
1-\left(1+\dfrac{\rhom}{kT}w(\mathbf{q})\right)S(\bq)+\dfrac{\nu}{2} \dfrac{q_z}{q^2}\dfrac{\partial S(\bq)}{\partial q_x}=0.
\end{equation}
At first sight, the above discussion seems totally irrelevant to the hard sphere system because the hard sphere potential does not have a Fourier transform. However, it has been shown \cite{SchVin02} that it is useful to view the hard sphere system as an ideal gas with the constraint that the pair distribution function
\begin{equation}\label{eqn:g}
g(\br)=1+\dfrac{1}{(2\pi)^3\overline{\rho}}\int_{\mathbf{q}\neq\boldsymbol{0}}d\bq(S(\bq)-1)e^{i\bq\cdot\br}
\end{equation} 
vanishes for distances $r=|\br|$ smaller than the range of the hard sphere interaction, that is the sphere diameter $R$. The constraint is imposed by introducing a Lagrange multiplier $\lambda(\mathbf{k})$ instead of $w(\mathbf{k})\rhom/kT$ in Eq.\eqref{eqn:steadystate} and \eqref{eqn:PhatSimpl}. \\Since the constraint holds only for $r<R$, $\lambda(\mathbf{k})$ is a Fourier transform of a function $\hat{\lambda}(\br)$ that vanishes for $r\geq R$. \\Thus, the full solution involves obtaining $S(\mathbf{q})$ as a functional of such a $\lambda(\mathbf{q})$ from the equation 
\begin{equation}\label{eqn:S}
1-(1+\lambda(\bq))S(\bq)+\dfrac{\nu}{2}\dfrac{q_z}{q^2}\dfrac{\partial S(\bq)}{\partial q_x}=0.
\end{equation}
Then $\lambda(\mathbf{q})$ has to be determined from the condition 
\begin{equation}\label{eqn:constraint}
g(\br)=0\quad\textrm{for }r<R.
\end{equation}
It is easy to see that in the absence of shear the conditions described by Eq.\eqref{eqn:g}, \eqref{eqn:S} (with $\nu=0$) and \eqref{eqn:constraint} and the condition on $\hat{\lambda}(\br)$ are equivalent to the hard sphere Percus-Yevick equation \cite{PY}.\\

The Percus-Yevick approximation leads to an analytic solution for $\lambda(q)$ which yields very good results for liquid densities of the system. Above those densities, though, the pressure predicted by that approximation shows no sign of termination of the liquid phase. \\
Fortunately, however, the pair distribution function, which is a probability distribution and therefore non-negative, vanishes first at some distance for a threshold value of the volume fraction, $\eta_c$. For higher volume fractions, the pair distribution function becomes negative for a range of distances, indicating the failure of the Percus-Yevick approximation at volume fractions higher than the threshold value \cite{schwartz} .\\

Since the vanishing of the pair distribution function at some distance indicates caging, it has been suggested in the above reference that the threshold volume fraction, although approximate, has a physical meaning of jamming or random close packing. This suggestion is supported by the comparison of the threshold volume fraction for dimensions between 3 to 9 obtained from the Percus-Yevick equation \cite{schwartz} and the corresponding values for random close packing (RCP) \cite{RCP} or maximal random jammed (MRJ) \cite{MRJ} systems.\\

In fact, this is one of the main motivations for the present work. It is well known that a jamming transition may occur in systems of objects subject to shear \cite{08Lu}. This jamming transition may be related, like in the absence of shear, to the vanishing of the pair distribution function. So, the study of the effect of shear on the pair distribution function is one of the main goals of the present article. 

Therefore, we turn our attention back to the general case of the finite shear, described by Eq.\eqref{eqn:g}-\eqref{eqn:constraint} and $\hat{\lambda}(\br)=0$ for $r>R$.

This is a formidable set of equations, much more complicated than the original hard sphere Percus-Yevick equation. We may, however, try to construct a solution as follows. Since the result for the hard sphere system without shear is known, we expect to be able to write our solution as an expansion in $\nu$ around the (known and exact) Percus-Yevick solution $(\lamPY(q),\SPY(q))$:
\begin{equation}
\begin{split}
&\lambda(\bq)=\sum_{i}\lambda_i(\bq)\nu^i,\\
&S(\bq)=\sum_{i}S_i(\bq)\nu^i,
\end{split}
\end{equation}
where $(\lambda_0(\bq),S_0(\bq))\equiv(\lamPY(q),\SPY(q))$ with $q=|\bq|$. This will result in $g(\br)=\sum_i g_i(\br)\nu^i$ with $g_0(\br)=g_{PY}(r)$. \\

In the present article, we will just concentrate on the first order expansion of $\lambda(\bq)$ and $S(\bq)$. Higher orders follow the same strategy that will be presented in the following but, although conceptually straightforward, are considerably more complicated. 

The zero order of the system is obviously satisfied by $(\lamPY(q),\SPY(q))$. From the first order in $\nu$ we obtain the following equation relating the couple $(S_1(\bq), \lambda_1(\bq))$,
\begin{equation}\label{eqn:unowithrotation}
-(1+\lamPY(q))S_1(\bq)-\SPY(q)\lambda_1(\bq)+\dfrac{1}{2}\frac{q_zq_x}{q^3}\frac{d \SPY(q)}{d q}=0.
\end{equation}
The form of the last term of the equation conveys the kind of dependence on the direction of $\bq$, so that we can write
\begin{equation}\label{eqn:l1S1}
\begin{split}
&\lambda_1(\bq)=\frac{q_xq_z}{q^2}f_1(q),\\
&S_1(\bq)=\frac{q_xq_z}{q^2}h_1(q).
\end{split}
\end{equation}
Eq.\eqref{eqn:unowithrotation} is then written for $f_1(q)$ and $h_1(q)$ and eventually provides the relation between these two quantities, which depend only on the absolute value of $\bq$:
\begin{equation}\label{eqn:relationSlambda}
h_1(q)=-\SPY^2(q)f_1(q)+\frac{1}{4q}\frac{d\SPY^2(q)}{d q}.
\end{equation}
The relation above gives $S_1(\bq)$ in terms of $\lambda_1(\bq)$, in analogy to the simple relation between $S_{PY}(q)$ and $\lambda_{PY}(q)$ in the absence of shear.\\

Now that we have reduced the problem to finding a function of a scalar value, we can turn to the evaluation of the pair distribution function as a functional of the unknown $f_1(q)$. This function will then be determined from condition \eqref{eqn:constraint} above.\\
It proves useful to express the directional dependence of the integrand on the right hand side of Eq.\eqref{eqn:g} in terms of spherical harmonics. We use the expansion of the plane wave in terms of spherical harmonics and the relation $q_xq_z/q^2=\sqrt{2\pi/15}\left[Y_2^{-1}(\theta_{\bq},\varphi_{\bq})-1/6Y_2^{1}(\theta_{\bq},\varphi_{\bq})\right]$. The angles $(\theta_{\bq}, \varphi_{\bq})$ determine the direction of $\bq$. \\
Eq.\eqref{eqn:g} results now in a single equation for the first order correction in $\nu$,
\begin{equation}\label{eqn:final}
\int_0^{+\infty} q^2\left[ \frac{1}{4q}\frac{d\SPY^2(q)}{d q}-\SPY^2(q)f_1(q)\right]j_2(qr)dq=0
\end{equation}
for $r<R$, where $j_2(qr)$ is the spherical Bessel function of order two.\\

Let us denote by $\hat{f}_1(r)$ and $\hat{\lambda}_1(\br)$ the functions whose Fourier transforms are $f_1(q)$ and $\lambda_1(\bq)$ respectively. 
We use a parametrization for $\hat{f}_1(r)$ that ensures that $\hat{\lambda}_1(\br)$ vanishes for $r>R$.\\
The relation between $\lambda_1(\bq)$ and $f_1(q)$, following Eq. \eqref{eqn:l1S1}, implies
\begin{equation}\label{eqn:l1_f1}
\nabla^2\hat{\lambda}_1(\br)=\dfrac{\partial^2}{\partial x\partial z}\hat{f}_1(r).
\end{equation}
By consequence, we obviously need to require that $\hat{f}_1(r)=0$ for $r>R$. But this is not the whole story: Eq.\eqref{eqn:l1_f1} can be solved to yield
\begin{equation}
\hat{\lambda}_1(\br)=-\dfrac{\partial^2}{\partial x\partial z}\phi(r),
\end{equation}
where the \emph{spherically symmetric potential} $\phi(r)$ is written in terms of the spherically symmetric \emph{charge density} $-\hat{f}_1(r)/4\pi$,
\begin{equation}
\phi(r)=-\dfrac{1}{4\pi}\int d\br'\dfrac{\hat{f}_1(r')}{|\br-\br'|}=\dfrac{Q}{r}.
\end{equation}
The last equality on the right hand side of the above equation holds for $r>R$ and $Q$ is the total \emph{charge} contained in a sphere of radius $R$. Thus, it follows that not only the charge density has to vanish outside the sphere, but also the total charge within the sphere has to be zero. This implies
\begin{equation}\label{eqn:Qeq0}
\int_0^R r^2\hat{f}_1(r)dr=0. 
\end{equation}
We use a parametrization for $\hat{f}_1(r)$ that ensures that $\lambda_1(\br)$ vanishes for $r>R$,
\begin{equation}
\hat{f}_1(r)=\left\{\begin{array}{ll}
\sum_{j=0}^n a_j r^j & \textrm{for } r\leq R\vspace{0.2cm}\\
0 & \textrm{elsewhere},
\end{array}\right.
\end{equation}
in terms of the coefficients $\{a_j\}$. 

Eq.\eqref{eqn:final} is used now to find the coefficients $\{a_j\}$. When doing that, we must recall that those coefficients are constrained by the requirement $Q=0$ (Eq.\eqref{eqn:Qeq0}).\\
Once $f_1(r)$ is known, we can easily set $S_1(\bq)$ and $g_1(\br)$ (practically, we choose $n=10$ independent unknowns $a_j$ and this ensures us that the root mean square of $g_1(\br)$ in the interval $(0,R)$ is of the order of $10^{-4}$).

Now, we can use the solution $(\lambda_1(\bq),S_1(\bq))$ to solve the equation for the second order term in $\nu$ but we postpone that to future study.
\begin{figure}%
\centering
\subfigure[$g_{PY}(r)$]{\includegraphics[width=.5\textwidth, clip, keepaspectratio]{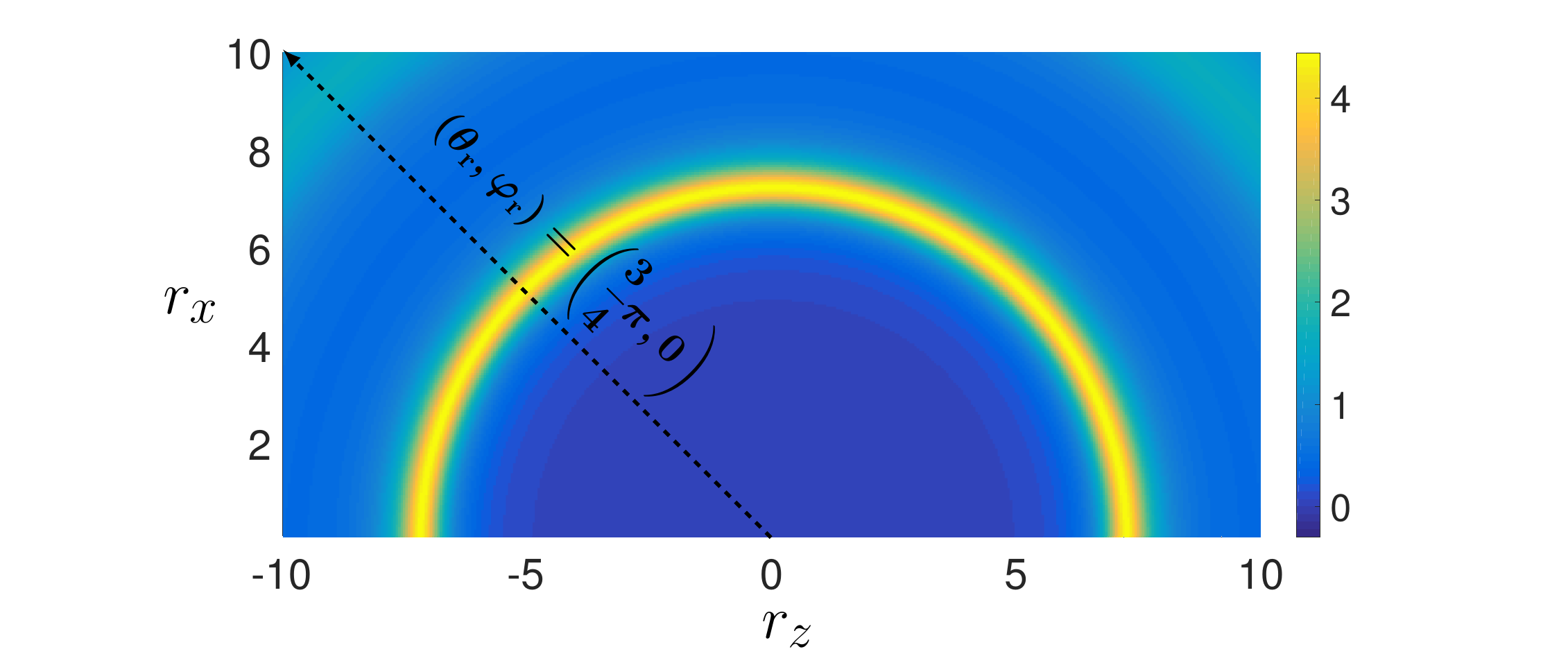}\label{fig:SPY}}
\subfigure[$g_1(\br)$]{\includegraphics[width=.5\textwidth, clip, keepaspectratio]{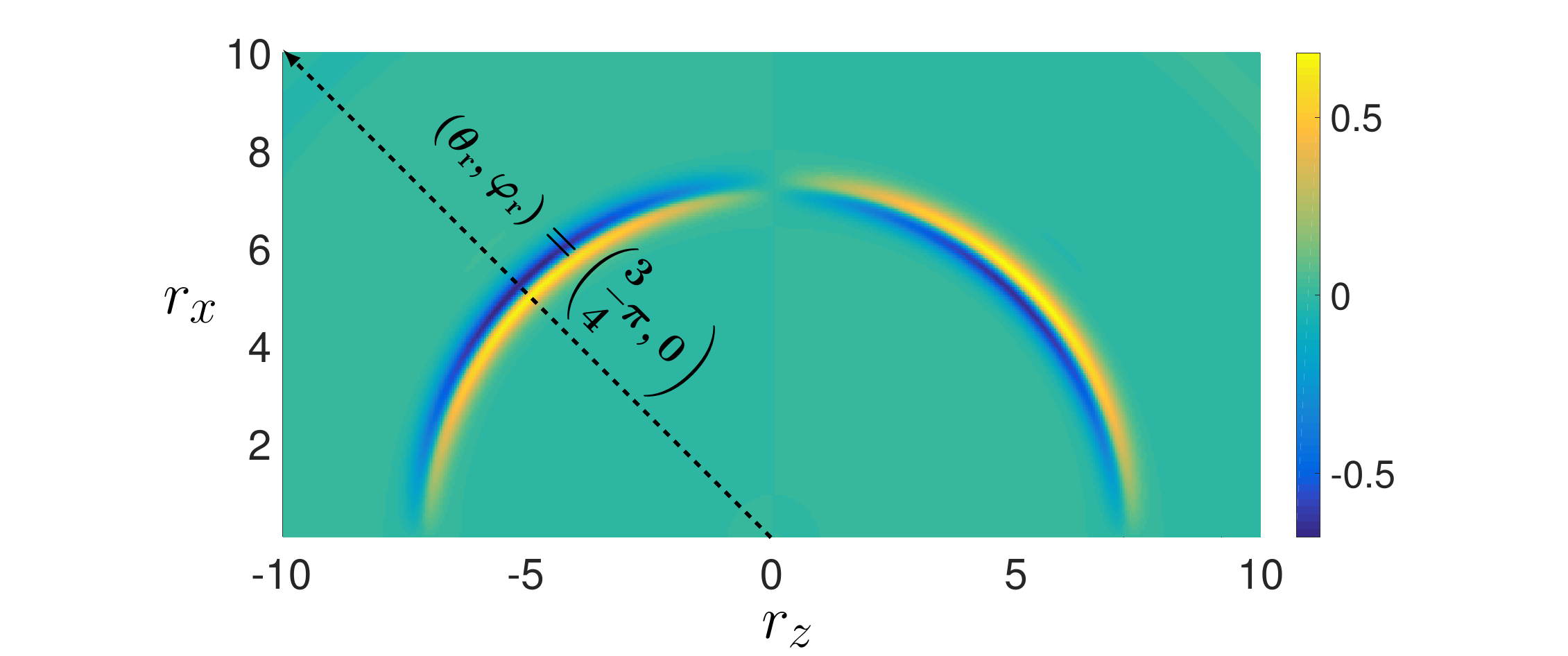}\label{fig:S1}}\\
\caption{The terms $g_{PY}(r)$ and $g_1(\br)$ in the $(r_x,r_z)$ plane, computed for $\eta=0.55$. The distance $r$ is given in units of $R$. The dashed line underlines the direction determined by the angles $(\theta_{\br}, \varphi_{\br})=\left(3\pi/4,0\right)$, whose importance will be explained later on.}
\label{fig:gofr_2d}
\end{figure} 
In order to work with dimensionless quantities, we use the volume fraction $\eta=\pi R^3\overline{\rho}/6$ and we define the \emph{P\'eclet number} $P=\nu R^2$. The pair distribution function is then expressed as $g(\br)=g_{PY}(r)+P/R^2\,g_1(\br)$. \\
Fig. \ref{fig:SPY} shows $g_{PY}(r)$ in the $(r_x,r_z)$ plane. Obviously, $g_{PY}(r)$ is just a function of $r_x^2+r_z^2$. Fig. \ref{fig:S1} shows $g_1(\br)$ in the same plane: it has, in addition, also an angular 
dependence. In both plots, a dashed line indicates the direction of $\br$ determined by the angles $(\theta_{\br},\varphi_{\br})=\left(3\pi/4,0\right)$.

\begin{figure}[htbp]
\begin{center}
\includegraphics[width=.5\textwidth, clip, keepaspectratio]{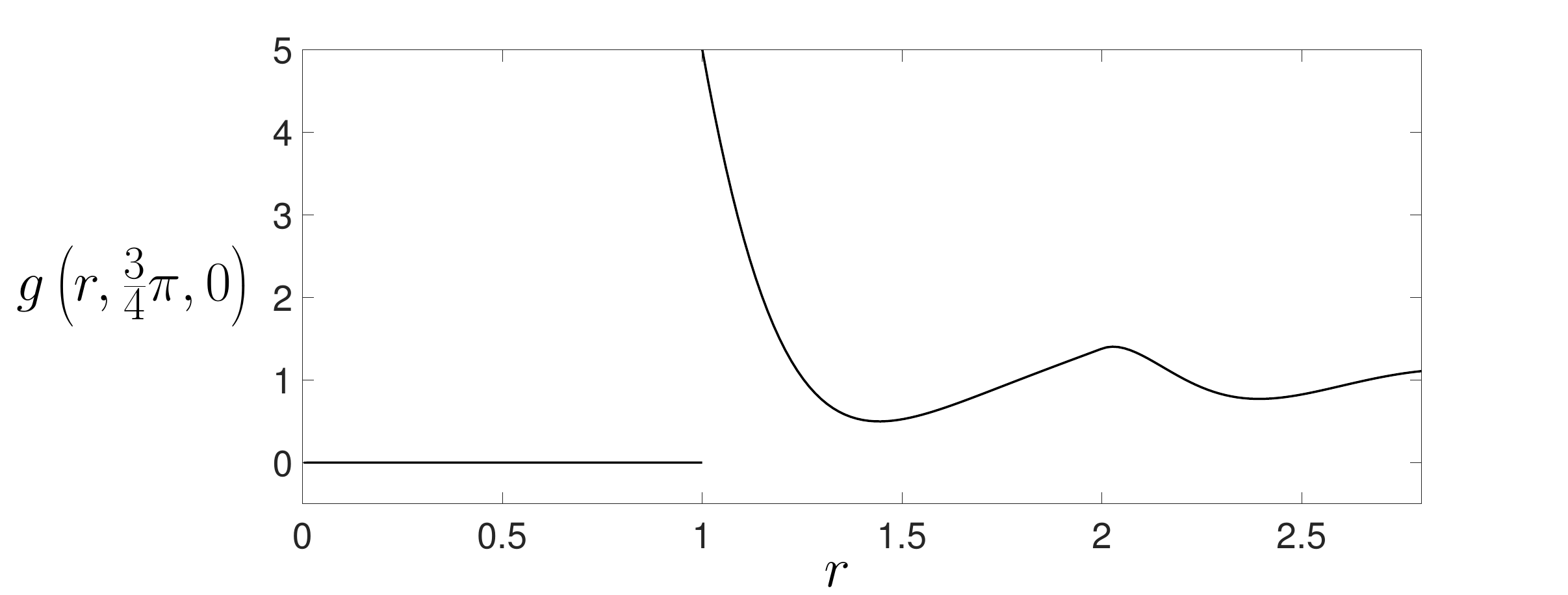}
\caption{The pair distribution function for $(\eta, P)=(0.5,1)$ in the direction determined by $(\theta_{\br}, \varphi_{\br})=(3\pi/4, 0)$, where the angles $(\theta_{\br}, \varphi_{\br})$ are the direction of $\br$. The distance $r$ is given in units of $R$.}
\label{fig:gofr}
\end{center}
\end{figure}

Fig. \ref{fig:gofr} shows the behavior of $g\left(r,3\pi/4,0\right)$, that is the pair correlation function exactly in this direction, for $\eta=0.5$ and $P=1$. 
The choice of the spatial angle is not accidental: it corresponds to the direction where $g_1(\br)$ attains its lowest value for $r$ close to the minimum point of $g_{PY}(r)$. The interest in this direction will be better explained in the following.

It is particularly interesting to notice that the positivity condition for $g(\br)$ is satisfied for the value of the couple $(\eta, P)$ in Fig. \ref{fig:gofr}.\\
As already said, this condition fails to be satisfied already in the case without shear for values of $\eta$ beyond a certain threshold and this event can be interpreted as occurrence of jamming in the suspension.  By adding external shear, we simply expect to facilitate the phenomenon of jamming for smaller $\eta$. Practically, we want to see when $g(\br)$ vanishes for the first time. This means, first of all, that the contribution of $g_1(\br)$ must be negative and as significant as possible. Second, this has to happen close to the minimum point for $g_{PY}(r)$. As Fig. \ref{fig:gofr_2d} shows, this happens for $(\theta_{\br}, \varphi_{\br})=(3\pi/4, 0)$.\\

\begin{figure}[htbp]
\begin{center}
\includegraphics[width=.5\textwidth, clip, keepaspectratio]{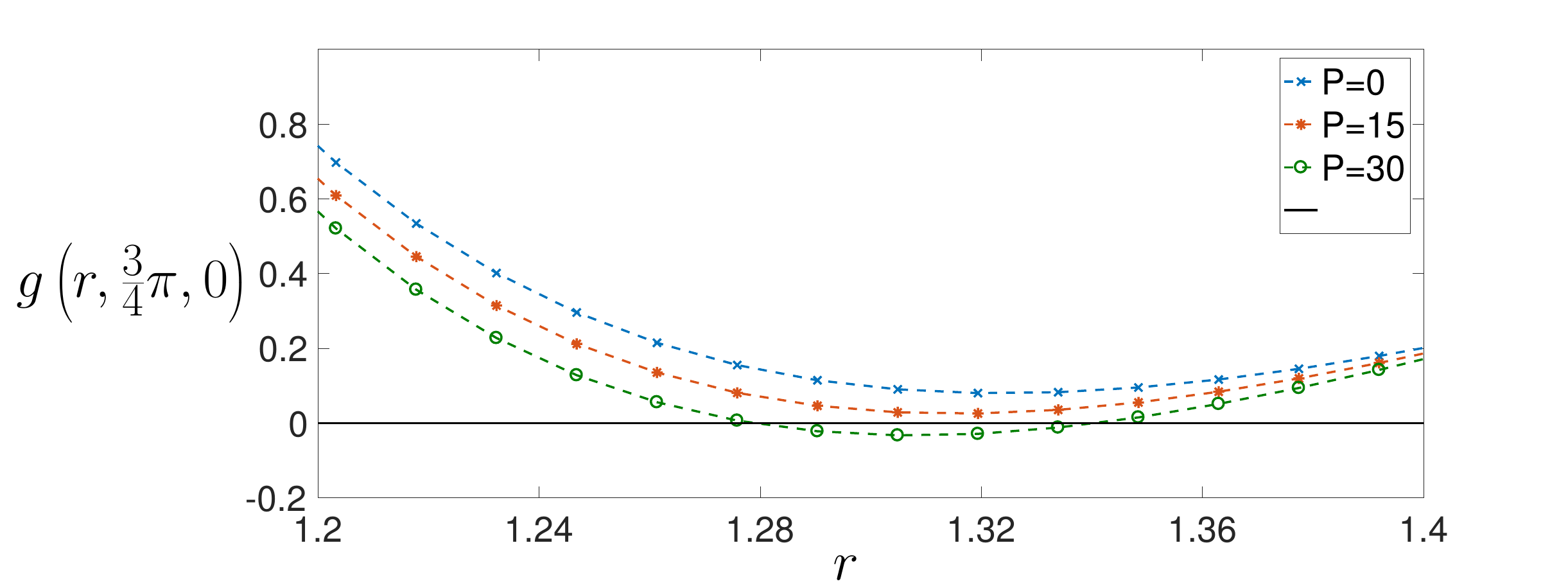}
\caption{The pair distribution function in the direction $(\theta, \varphi)=(3\pi/4, 0)$ for $\eta=0.6$ and different values of $P=0,15,30$.}
\label{fig:gofr_vs_Pe}
\end{center}
\end{figure}

Let us choose a fixed volume fraction and vary the shear. Fig. \ref{fig:gofr_vs_Pe} shows the behavior of $g(r, 3\pi/4,0)$ around its minimum point for $\eta=0.6$ and different values of the P\'eclet number. For small $P$ the curve is always positive but it becomes negative for increasing values of $P$, as expected. \\

\begin{figure}[htbp]
\begin{center}
\includegraphics[width=.5\textwidth, clip, keepaspectratio]{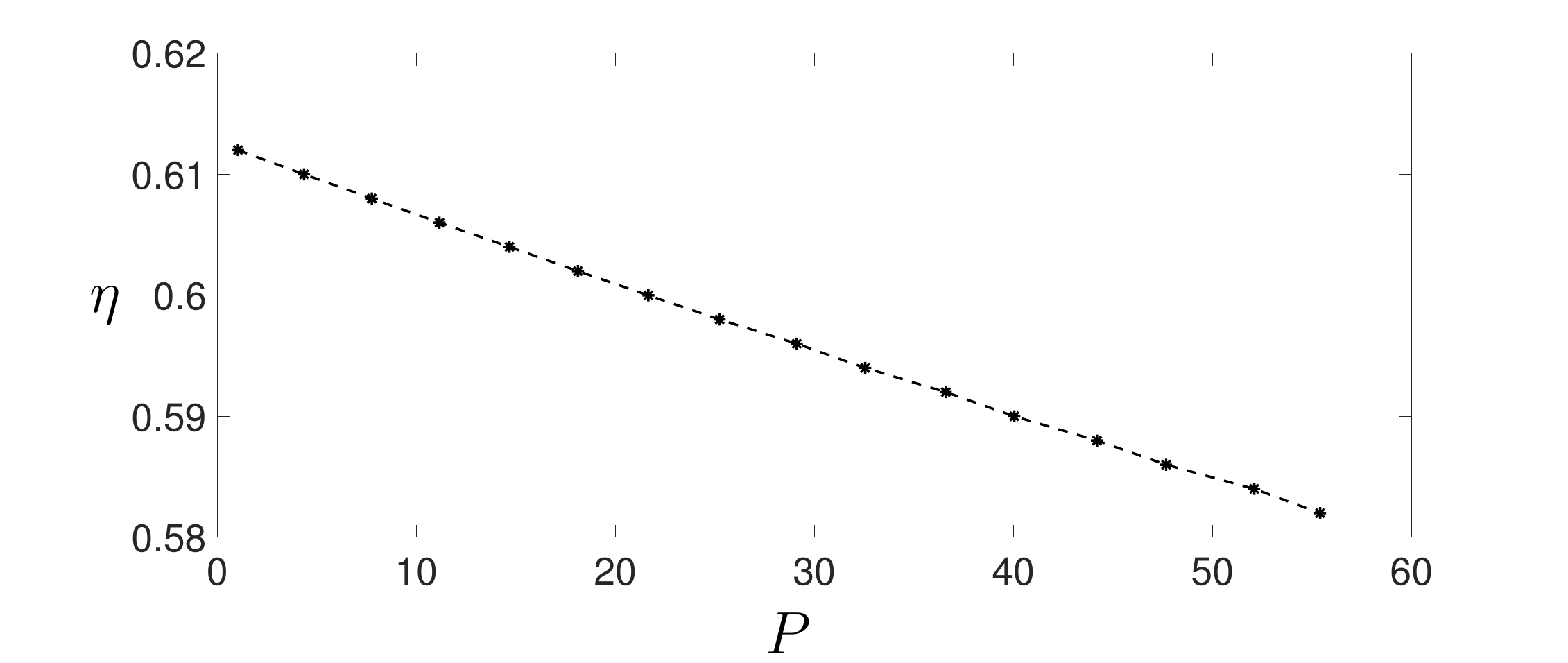}
\caption{Border between the fluid phase and the jammed phase in the $(P,\eta)$ plane.}
\label{fig:eta_vs_Pe}
\end{center}
\end{figure}

This leads to our final result, that is a phase diagram in the plane $(P, \eta)$ as shown in Fig. \ref{fig:eta_vs_Pe}. \\
The curve can be calculated for bigger values of $P$ but it shows a slight deviation from the linear behavior that has to be expected within the first order approximation. 

ne
In conclusion, by combining previous analytic results for two primary problems, it is possible to write a system of equations for hard sphere suspensions subject to external shear. \\

We propose a suitable form of the solution, as an expansion in the shear rate around the Percus-Yevick solution for non-sheared hard sphere suspension. This allows us to capture the distortion of the structure factor and its angular dependence for small shear. From this distortion we obtain the (deformed) pair distribution function and the volume fraction for which it vanishes for the first time.\\
Consequently, we derive the border between the fluid phase and the jammed phase in the suspension, as a function of volume fraction and shear.

The problem has still many aspects to investigate. A natural next step is the comparison with numerical and experimental results.\\
Moreover, a better approximation for the solution, hopefully valid for a wider range of shear rates, will be the object of future research.\\

\end{document}